\documentclass[%
 reprint,
preprint,
 amsmath,amssymb,
 aps,
pra,
onecolumn,
]{revtex4-2}

\usepackage{xcolor}
\usepackage{subcaption}
\usepackage{graphicx}
\usepackage{dcolumn}
\usepackage{bm}

\newcommand{\ket}[1]{|#1\rangle}
\newcommand{\bra}[1]{\langle #1|}

\begin{document}


\title{In situ calibration of unitary operations during quantum error correction}

\author{Jonathan Kunjummen}
\affiliation{%
Joint Center for Quantum Information and Computer Science, University of Maryland-NIST, College Park, Maryland 20742, USA}
\affiliation{Joint Quantum Institute, University of Maryland-NIST, College Park, Maryland 20742, USA}
\author{J. M. Taylor}%
 \email{jmtaylor@umd.edu; current address: Axiomatic AI, Cambridge, MA 02139, USA}
\affiliation{%
Joint Center for Quantum Information and Computer Science, University of Maryland-NIST, College Park, Maryland 20742, USA}
\affiliation{Joint Quantum Institute, University of Maryland-NIST, College Park, Maryland 20742, USA}

\date{\today}

\begin{abstract}
Quantum error correction uses the measurement of syndromes and classical decoding algorithms to estimate the location and type of errors while protecting the encoded quantum bits. Here we consider how prior information and Bayesian updates can play a critical role in improving the performance of QEC in the scenario of a particularly noisy qubit. This allows for leveraging even distance codes, which typically are less valued in QEC, to handle the noisy qubit, changing the power-law scaling of the logical error rate with the baseline physical error rate. A crucial component of this is updating the prior by real time feeding of decoder outputs into a approximate Kalman filter. Thus our approach provides a bootstrap to the actual error rates. We show this via simulation of the full closed-loop system: starting from uniform priors, the update procedure gradually learns site-specific error rates, enabling the decoder to outperform a fixed-prior baseline. In turn, we show that this enables in situ calibration of unitary operations via injection of gate set tomography operations with only moderate overhead in the more typical scenario of low noise qubits. 
\end{abstract}

\maketitle


\section{\label{sec:intro}Introduction}

A long-standing goal central to realizing the power of quantum computation in practice is the creation and execution of a fault-tolerant, error-corrected quantum computer. A fault-tolerant machine works in theory under key provisions: uncorrelated, static, well-calibrated errors. With practical quantum error correction now being demonstrated experimentally \cite{google,harvard,atomcomputing,quantinuum,tesseract}, a number of  different groups have studied adapting fault-tolerant principles to practical situations:  calibration errors, correlated errors in space and time, and broken qubits and couplers \cite{luci,SnL}.  Alongside progress in more realistic error models, theoretical progress has been made on tailoring error correction to structured or biased noise \cite{xzzx, spin-biased-noise,tailored-small-codes}, or on the other hand developing protocols to restructure noise through techniques like randomized compiling \cite{randomized-compiling,rc-in-qec}. In addition improvements in fast and accurate decoders which with finite information processing capability have them now approaching the theoretical limits on perfect decoders. Nevertheless, a variety of key questions remain in how we can leverage advances to improve performances of devices under realistic noise models.

This work contributes to these developments by investigating how prior knowledge of device physics can be usefully deployed in the decoding context. We begin with a theoretical result providing guarantees from code distance on the ability to correct errors in a mixture of known and unknown locations. We show the practical consequences of this formal result by dynamically adapt the Belief Propagation decoder to slowly changing error models. 

Furthermore, we use our adaptive decoder system to implement a new protocol, in which we inject controlled rotations that have a slight over- or under-rotation that we want to estimate, and use decoder outcomes to estimate process information. This allows in situ update of calibration operation while also performing error correction. We find that the theoretically expected outcome of known locations and unknown locations corresponds to our observed numerical results, i.e., that even-distance codes allow one to correct one introduced error of known location relative to odd-distance of one distance smaller. Our results are only shown for phenomenological noise: in this work we assume perfect syndrome extraction. We anticipate that a scalable extension of this calibration to the case of circuit level noise is possible as the influence of the error on the syndrome extraction circuit can be explicitly protected against via dynamic modification of the circuit to prevent hook errors, but leave this for future work. 

It is instructive to compare this work to the contemporary literature on error correction with erasure errors \cite{erasure-atom-loss, erasure-ft-architecture, erasure-magic, erasure-minesweeper, erasure-optimization, erasure-original, erasure-Yb171}. Erasure error correction also has the attractive feature that the errors considered are of known location and therefore can be corrected up to higher weight than in the unknown case. This techniques have been developed with particular attention to superconducting and neutral atom platforms, where erasures are heralded through leakage to levels outside the qubit subspace or to atom loss. In contrast to to the erasure literature, our approach is more situated around solid state paradigms where qubits drift out of calibration.

\section{Theory of correcting errors in a mixture of known and unknown locations}

We begin by reviewing and then extending with a small lemma the capabilities of quantum error-correcting codes for the case of a mixture of known and unknown error locations. Recall that the code distance $d$ is defined by the minimum weight of a Pauli string $P$ on the code that provides some logical operation, i.e. 
\begin{equation}\label{eq:distance}
        d \equiv \min \text{wt}\, P ,\;\bra{i_L} P \ket{j_L} \neq c_P \,\delta_{ij}
\end{equation}
for some logical basis states $\ket{i_L}, \ket{j_L}$ of the code and for $c_P$ some constant independent of the logical states. The definition of distance is related and in fact complementary to the Knill-Laflamme conditions for error correction \cite{kl-conditions}, which state that a set of errors $\mathcal{E}$ is correctable by a code if, for every $E_a, E_b \in \mathcal{E}$, and every pair of logical basis states as above,
\begin{equation}\label{eq:KL-conditions}
    \bra{i_L} E_b^\dag E_a \ket{j_L} = c_{ab} \,\delta_{ij}
\end{equation}
with $c_{ab}$ again a constant independent of $i,j$. In this proof (and throughout this work) we assume perfect stabilizer measurement and perfect implementation of correction operations; we anticipate that future work will handle more realistic noise models. Consistent with the field, we only consider error operations that are Pauli strings. There are a number of fundamental properties that are guaranteed by having a certain code distance, which we shall review, in order to set up a subtlety in comparing even- and odd-distance codes which is the key theoretical insight for this work. 

First, let us review the typical statements of QEC performance:

\textbf{Statement 1} (correcting errors of known location): If the support of errors is limited to $d-1$ physical qubits, up to $d-1$ errors can be corrected.

Proof: Let $\mathcal{E}$ be the set of all errors supported on a known set of $d-1$ physical qubits. Then for any $E_a, E_b \in \mathcal{E}$,  $E^\dag_b E_a$ is supported on that same set of $d-1$ qubits, and so $\text{wt}(E^\dag_b E_a) < d$. As a result of the definition of distance in Eq.\ref{eq:distance},  $E_a$ and $E_b$ must satisfy the Knill-Laflamme conditions of Eq.\ref{eq:KL-conditions}, and so the entire set $\mathcal{E}$ is correctable.

\textbf{Statement 2} (correcting errors of unknown location): Up to $\lfloor \frac{d-1}{2} \rfloor$ errors of unknown location can be detected and corrected. 

Proof: Let $\mathcal{E}$ be the set of all errors of weight at most $\lfloor \frac{d-1}{2} \rfloor$. Now for any $E_a, E_b \in \mathcal{E}$ we have $\text{wt} (E_b^\dag E_a) \leq \text{wt} (E_b) + \text{wt} ( E_a) < d$, and from here the logic is identical to the preceding case.

Critically, the decoding for statement 1 takes advantage of the knowledge of the locations, while statement 2 does not. This leads to a very different performance -- $d-1$ vs $\lfloor \frac{d-1}{2} \rfloor$. As is well known, for the task of Statement 2, even-distance codes offer no advantage over codes with distance one unit smaller. However, there clearly is a difference for the case of Statement 1. With this in mind, we suggest there should be a formal scenario in which, when particular qubits have much higher error rates than the majority, even codes can outperform odd ones in correcting errors in a mixture of arbitrary and preferred locations. We prove this with the following lemma:

\textbf{Lemma} (even-odd separation for detection and correction with a preferred site): Given a code of distance $d$ and knowledge that errors occurred on a set of $n_1<d$ sites, a code can additionally correct up to $n_2$ errors of unknown location with $n_1 +2 n_2 < d$.

Proof: Let $\mathcal{E}$ be the set of all errors with arbitrary support on the known $n_1$ sites and support of size at most $n_2$ on the remaining sites. Then for any $E_a, E_b \in \mathcal{E}$, the product $E^\dag_b E_a$ is supported on at most $2 n_2$ sites apart from the known $n_1$ sites, and so the weight of $E^\dag_b E_a$ can be bounded as $\text{wt}(E^\dag_b E_a) \leq n_1 + 2n_2 < d$, from which the correctability of $\mathcal{E}$ follows as before.

One particular consequence of this result is that while an even-distance code cannot correct any more errors in unknown locations than an odd-distance code of distance one unit less, the even code can correct \textit{one additional error} of known location in the presence of the same number of unknown errors. Our result also allows for an intuitive explanation of the correction strategy. For stabilizer codes, since any pauli string squares to the identity, one can choose the set of correction operations to be the same as the set of correctable errors. The correction strategy in the scenario we examine, with a mixture of $n_2$ errors in known and $n_1$ errors in unknown locations, is that high-weight (weight more than $n_2$) corrections are allowed only if their additional support is on the sites corresponding to the known error locations. We now show practically how to realize this benefit in a much less restricted setting, in which a single qubit has a high error rate, by adapting the decoder to the task.

\begin{figure}
    \centering
    \includegraphics[width=0.9\linewidth]{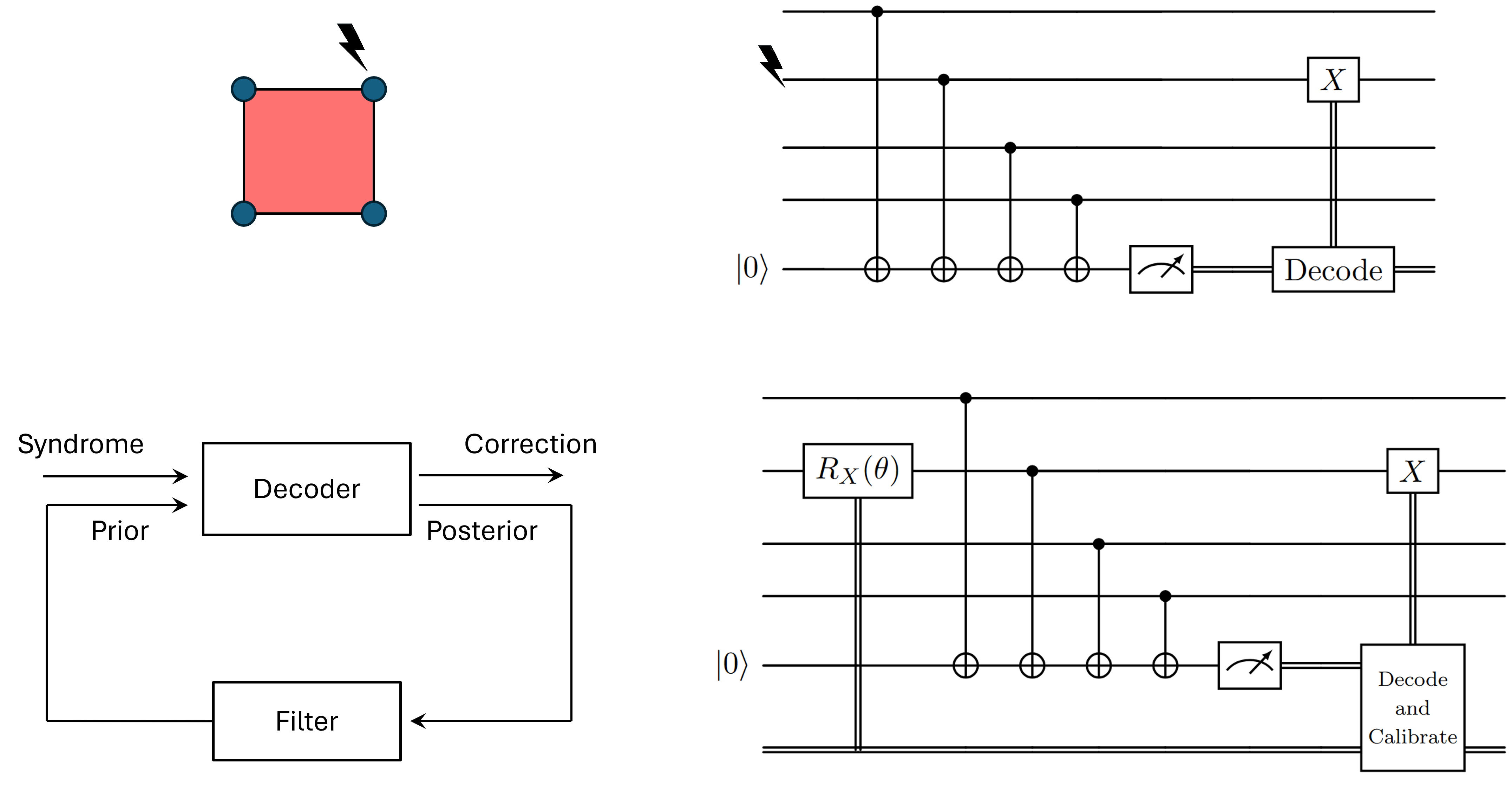}
    \caption{The intuition behind this work, for the case of the toy example: the [[4,2,2]] code. (Top left) The [[4,2,2]] code, composed of four data qubits (blue dots) has a $Z$-stabilizer (central square) which touches all of them. The top right qubit is especially error-prone leading to a bit flip error  (lightning), stabilizer measurement registers an error (red). (Bottom left) The control flow diagram of this work. A decoder using prior error rate information can correct the error shown above without inducing a logical error, and the decoding result can then be used to update the prior. (Right) Circuits showing application of prior-assisted decoding to improve error correction performance (top) by preferring corrections where priors are high, and to in situ gate calibration (bottom) by inferring the value of gate parameters from decoding data while protecting logical information.}
    \label{fig:placeholder}
\end{figure}

\section{Site-specific information in BPOSD}

In the previous section we proved that the artificial scenario in which, e.g., one physical qubit in a code is taken to have a unit error rate, moderate code distance increases (e.g., odd to even) yield potentially large improvements in performance. The goal of the current section is to show that this formal observation has practical consequences. As a first step toward increasing the sophistication of decoding to take advantage of this realization, we implement a decoder that demonstrates the notionally obvious statement that syndrome information plus prior information about high rate error locations can lead to better decoding than syndrome information alone.

We focus specifically on the case of CSS codes, which allow us to leverage existing classical decoders without substantial modification as their X and Z check matrices can be processed separately. As a reminder, a decoding algorithm, when successful, takes a set of stabilizer measurement outcomes (a syndrome) $s$ and outputs a correction operation $e$ that is stabilizer-equivalent to the error which occurred, i.e. the product of the correction operation and the measured error is stabilizer equivalent to the identity, correcting the error. A CSS code with $m_z$ Z-stabilizers and $m_x$ X-stabilizers defined by the parity check matrices $H_z, H_x$  \cite{calderbank-shor,steane}. Because the decoding problems for $X$ and $Z$ errors are independent for CSS codes, in the work that follows we shall always focus on a single category of errors (here, bit flips). In the numerical experiments to follow, the parity check matrix $H$ is always $H_z$ for the given code, and the error $e$ is always taken to be a product of bit flips.

In order to use a relatively standard decoder for our problem where one qubit has a high error rate, we leverage an adaptable decoder system that takes advantage of a prior expected distribution of error probabilities. While there are a variety of choices, here we use the Belief Propagation with Ordered Statistics decoding (BPOSD) algorithm, a popular decoder for quantum low-density parity check (qLDPC) codes that is based on the message-passing belief propagation algorithm from Bayesian inference on graphical models \cite{panteleev-bposd}. A key feature of this decoder for our work is that it incorporates prior estimates of the error rate for each qubit. As we will see later, we can also update these estimates after each successful decoding round to adapt the decoder to changing qubit error rates, which allows the decoder to adapt to `noisy' qubits over time.

The BPOSD algorithm is defined in terms of the code's Tanner graph, a bipartite graph with one set of vertices for the physical qubits, the second set for the parity checks, and the adjacency matrix between the two sets given by the parity check matrix itself. Beginning with prior estimates of the error rates for each qubit, the belief propagation part of the algorithm iteratively updates the posterior estimated probability that an error occurred on each qubit, conditioned on the observed syndrome and the (iteratively updated) error probabilities of neighboring qubits in the Tanner graph. The belief propagation algorithm is guaranteed to converge for tree-like graphs, but there is a fundamental ambiguity in quantum codes which can cause the algorithm to fail to converge, namely that the decoder cannot distinguish between errors which are stabilizer equivalent and of equal weight. The ordered statistics step takes over if belief propagation fails to converge to an error consistent with the observed syndrome. In its most basic form, OSD chooses the highest likelihood subset of error locations such that the syndrome equation $H \cdot e = s$ (where all arithmetic is carried out in $\mathbb{Z}_2$) is well-posed/invertible and then solves this equation.

\begin{figure}[htbp]
    \centering
    \includegraphics[width=\linewidth]{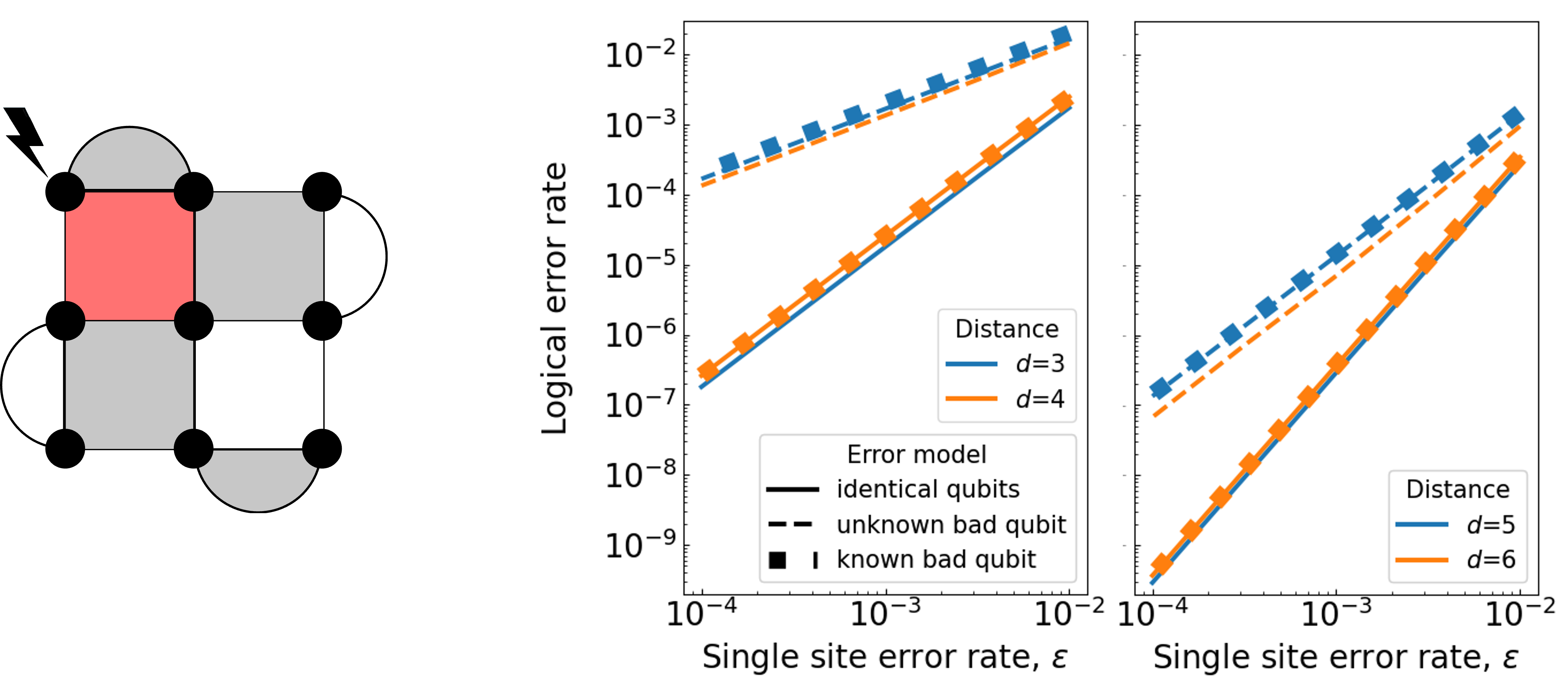}
    \caption{(Left) The rotated surface code. Data qubits are vertices of the graph whereas $X$-checks ($Z$-checks) are on the grey (white) faces. (Right) Logical error rate for $d=3$ up to $d=6$ rotated surface codes for each of the three cases from the main text, which we plot on a log-log scale versus  $\varepsilon$ to extract the scaling of logical state preservation with system-wide error rate. For $d=3$ and $d=4$,  the logical error rate for case 1 (case 2) has the same scaling for both distances, i.e. as $\varepsilon^2$ ($\varepsilon^1$), but case 3, however, $d=4$ scales better than $d=3$, as we expect from Statement 4. Specifically, the case 3 and case 1 lines are nearly indistinguishable for $d=4$ -- the code handles both an unknown and known error with $\varepsilon^2$ logical error. Analogous behavior holds for $d=5$ and $d=6$, but the scaling of all error rates increased by a power of $\varepsilon$.}
    \label{fig:smart-decoding}
\end{figure}

As described above, for a fixed code the inputs to the BPOSD decoder are the observed syndrome $s$ and a vector of estimated bit flip error rates for each qubit,
\begin{equation}
    \vec p_{flip} = \begin{pmatrix}
        p_1 & \cdots & p_N
    \end{pmatrix}^\intercal\ ,
\end{equation}
we are interested in the case where an isolated number of the qubits have error rates of order 1 while the others have some much smaller error rate $\varepsilon \ll 1$. In the numerics below we will focus on the case of a single bad qubit with bit flip rate $p_* \gg \varepsilon$
\begin{equation}
    p_i = \begin{cases}
        p_* & i = 1\\
        \varepsilon & i \neq 1
    \end{cases}\ .
\end{equation}
In Fig.~\ref{fig:smart-decoding} we show the effect of providing information on error-prone qubits to belief propagation. At a basic level, flagging a high-error-rate qubit ``breaks the tie'' between two complementary errors of the same size (complementary means they add up to a logical operation).
 
We carry out a numerical experiment on the rotated surface code. There are two models used in Fig~\ref{fig:smart-decoding}. As a control, we simulate a rotated surface code with perfect syndrome measurements where all qubits have identical and independent probabilities of error $\varepsilon$, which we vary. To investigate the effect of introducing a single error-prone site, possibly known and possibly not, we fix one bad qubit to have an error rate of bit flipping of $p_* = 1/3$, while the rest continue to have independent, identically distributed (i.i.d.) bit flips with probability $\varepsilon$. We remark here that BPOSD performance degrades as $p_* \rightarrow 1/2$, which is why we choose 1/3. 

For small distances ($d=3$ and $d=4$), we simulate the performance of a basic quantum memory experiment as follows. We calculate the probability of each of the $2^N$ possible bit flips that could happen, we pass the syndromes corresponding to each of these $2^N$ errors to the decoder. With the corresponding correction suggested by the decoder we then determine whether the final state is in the correct eigenstate of $Z_L$. Given a probability of each bit flip error and a success or failure outcome for each bit flip error, we can state the total failure probability of the quantum memory experiment. We examine three cases: 
\begin{enumerate}
    \item[Case 1:] The traditional scenario in which all qubits have identical bit flip probabilities and these are used to initialize the decoder;  (i.i.d. decoder with i.i.d. qubits, or `identical qubits')
    \item[Case 2:] Using the same initialization of the first case but having one ``bad'' qubit whose actual bit flip rate is close to 1/3; (i.i.d. decoder with bad qubit, or `unknown bad qubit') 
    \item[Case 3:] combining the error model of the second case with initialization of the decoder using that error model. (adapted decoder with bad qubit, or `known bad qubit')
\end{enumerate}


Fig.~\ref{fig:smart-decoding} displays the failure rates for the various models for a rotated surface code from distance $d=3$ up to $d=6$. In the case of a distance 4 code, the examined cases follow one of two scalings, either $O(\varepsilon)$ or $O(\varepsilon^2)$. This can be understood via a simple counting argument. In order to get an uncorrectable error we need an error of at least weight $(d-1)/2$. For $d=4$ this is a weight of $ \lceil 3/2 \rceil = 2$. This happens with probability $\varepsilon^2$ for min weight decoder-confounding errors which do not involve the bad qubit, and probability $p_* \varepsilon$ for errors which do involve the bad qubit. Importantly, if we do not pass information about the bad qubit to the decoder, we fail to achieve this scaling. 

Intuitively, passing these priors to the decoder has the result that whenever the decoder has the ability to choose between two equal-weight errors to explain the syndrome where one of those errors involves the bad qubit and one does not, it will choose the one which involves the bad qubit. This is consistent with our more formal proof, above, which treats corrections that include the known error location differently than those that do not. We note, too, that although a different approach to leveraging properties of even codes takes a decoded weight $d/2$ error as a trigger for postselection \cite{d4-postselection}, for us the difference in scaling achieved in the case of the known bad qubit does not require any postselection. The final panel of Fig.~\ref{fig:smart-decoding} shows that the change in scaling holds also for $d=5$ and $d=6$, with the logical error rate for the adapted decoder with bad qubit scaling as $\varepsilon^2$ for $d=5$ and as $\varepsilon^3$ for $d=6$.
For distances 5 and 6, to make the simulation more efficient we run the decoder only on errors of weight 6 or lower. The aggregate probability of a higher weight error is then calculated for each model and single site error rate; these are added to the plot as error bars and are too small to be seen.


According to our lemma, we expect that the tradeoff between errors of unknown location and known ones can be leveraged so that for every two error sites flagged with a high error rate, the number of errors on unknown locations that can be corrected goes down by one. However, we also believe that the formal statement of the tradeoff is overly pessimistic for practical settings, as it captures the worst-case scenario of nearby error-prone sites: sites which live on the same low-weight representative of some logical operator. 

Consider the surface code where two high-error sites live on opposite corners of the lattice. The shortest logical operator involving these two sites has weight scaling not as $d$ but as $2d$. If the error-prone sites are sparse in the sense that none of them are connected by low-weight logical operators, the errors which cause the decoder to fail will mostly likely be caused by a cluster of failures near a single error-prone site.  We conjecture in general that for randomly distributed defects in the surface code with number density $\nu$ that the failure rate $f$ depends on the generic error rate as $\log(f) \approx  (\frac{d}{2}-d\nu)\log(\varepsilon)$. This has the intuitive interpretation that what matters for the rotated surface code is the expected number of error-prone sites \textit{per row or column}, not the total.

\section{Filtering to learn error rates}
So far we have shown how decoders when primed with the right prior substantial outperform. In the examples of the previous step, we did not specify where the initial error rate estimates came from, and once set they were held fixed. Our next step is to show that the prior can be updated to a posterior, which in turn is passed to the next decoding round. Colloquially, since the output of a decoder is to estimate where bit flips occurred, this gives rise to time series data which can successfully update the prior error rates fed into the decoder.

Our update rule for learning bit flip rates from decoding data has the form of a Kalman filter with no applied control. Kalman filters are the optimal filter for linear systems; while we do not in general assume that error rates change according to this scenario, this should be correct for low frequency changes in qubit performance. 

To implement this, we simulate running a number of integer timesteps $t$, where the in cycle $t$ we prepare a logical state, idle, measure syndromes, and decode. We then update the qubit-wise error rates used in the BPOSD in time as 
\begin{equation}\label{eq:Kalman_update}
    p_{i, t+1} = p_{i,t} + \gamma \,(b_{i,t} - p_{i,t} ) 
\end{equation}
where $p_i$ is the prior estimate of the bit flip rate, $\gamma$ is the Kalman gain coefficient, and $b_i$ is the decoder output for the final probability of a bit flip at location $i$. Note that in this work we do not calculate the optimal Kalman gain, instead taking a fixed value; in practical settings, implementing a gain update would be appropriate as variances may change over time as well. 
Specifically, in the case of linear systems with Gaussian noise, where the Kalman filter is provably optimal, the gain is ultimately determined in the long time limit by the process and observation noise covariances. In our case these noise covariances are not only unknown, but the single-shot update $b_{i,t}$ is only an indirect estimate of the bit flip error rate on $i$, which is a nonlinear inference process of the decoding algorithm that goes from syndromes to an estimate of the qubit bit flips. Nonetheless, as we show below, this simple filter suffices.

We assume qubit behavior changes occur over long time scales, consistent with, e.g., $1/f$ noise. As a result, if we keep track of the history of decoder outputs, i.e., the estimated errors on each qubit, we can update our estimates of bit flip rates with the decoder output, giving future decoding instances information about which locations are likely to be noisy. In the case of static bit flip rates and perfect decoding, we would in general expect that the estimated bit flip rates converge to the true ones in time $\gamma^{-1}$ and to accuracy $\sqrt{\gamma}$.

\begin{figure}[htbp]
  \centering
  \includegraphics[width=\linewidth]{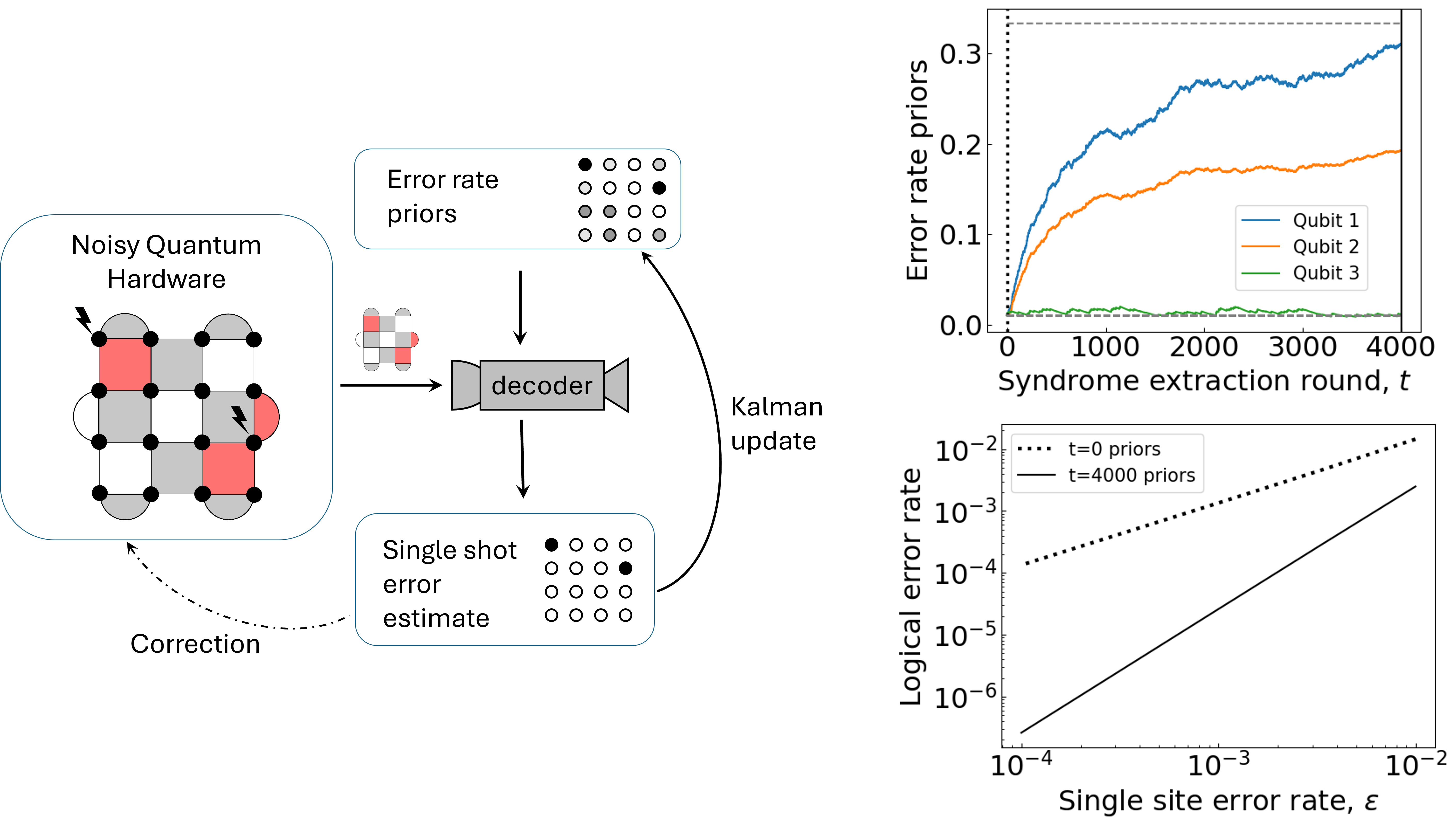}
  \caption{Application of the Kalman update on a distance 4 rotated surface code showing the enhanced performance after updating. (Left) The workflow: noisy quantum hardware gives rise to syndrome measurements, then the decoder decodes the syndrome and uses the resulting error estimate to update its priors and, possibly, apply a correction. (Top right) We plot the estimated bit-flip probabilities updated according to Eq.~\ref{eq:Kalman_update}. We see that the estimated rate for qubit 1 increases as it should, and qubit 2 does as well. (Bottom right) We then compare the performance of the decoder running on the priors at the beginning and end of the protocol and see different scaling with $\varepsilon$.}
  \label{fig:Kalman}
\end{figure}

Beginning from priors which assume that all qubits have the same error rate, we show for the
rotated surface code that our filtering decoder algorithm can self-consistently update, converging to the actual error rates. The results of the numerical experiment, shown in Fig.~\ref{fig:Kalman}, show the successful qualitative learning of the error rates. Starting from a prior in which all qubits are assumed to fail with the same small probability (scenario 1 from the previous section), the update from decoding data succeeds in increasing the estimated bit flip rate on qubit 1 to an order 1 probability. The error rate prior on qubit 2 also increases, a consequence of the fact that a bit flip on qubit 1 is stabilizer-equivalent to a bit flip on qubit 2 for the rotated surface code.

It is interesting to note that the estimated error rate oscillates around a level smaller than the true value, and in general the protocol does tends even in the long time limit to underestimate large error rates. This arises due to our choice of BPOSD. Specifically, the algorithm needs enough of a hint about relative likelihoods to get to the point where it is rounding correctly. So in many cases where the real answer was 0 or 1 the BP algorithm is turning that order 1 difference in behavior to a .4 vs .6 difference, thus underestimating the true variance. For example, we find that for the distance four codes, the average soft probability which gets rounded to a 1 is approximately 0.7. While soft probabilities are useful (they allow us to update our priors in a smooth manner) they also do better when the overall error probability is below 1/2. This observation is why we focus on error probabilities of 1/3 for the noisy qubit rather than 1/2, and accordingly in our calibration section, work away from the optimal Ramsey point. Regardless, as shown in Fig.~\ref{fig:Kalman}, the scaling performance remains $O(\varepsilon^2)$, consistent with successful application of our result.

\section{Application to gate calibration}

\begin{figure}[htbp]
\centering
     \includegraphics[width=\linewidth]{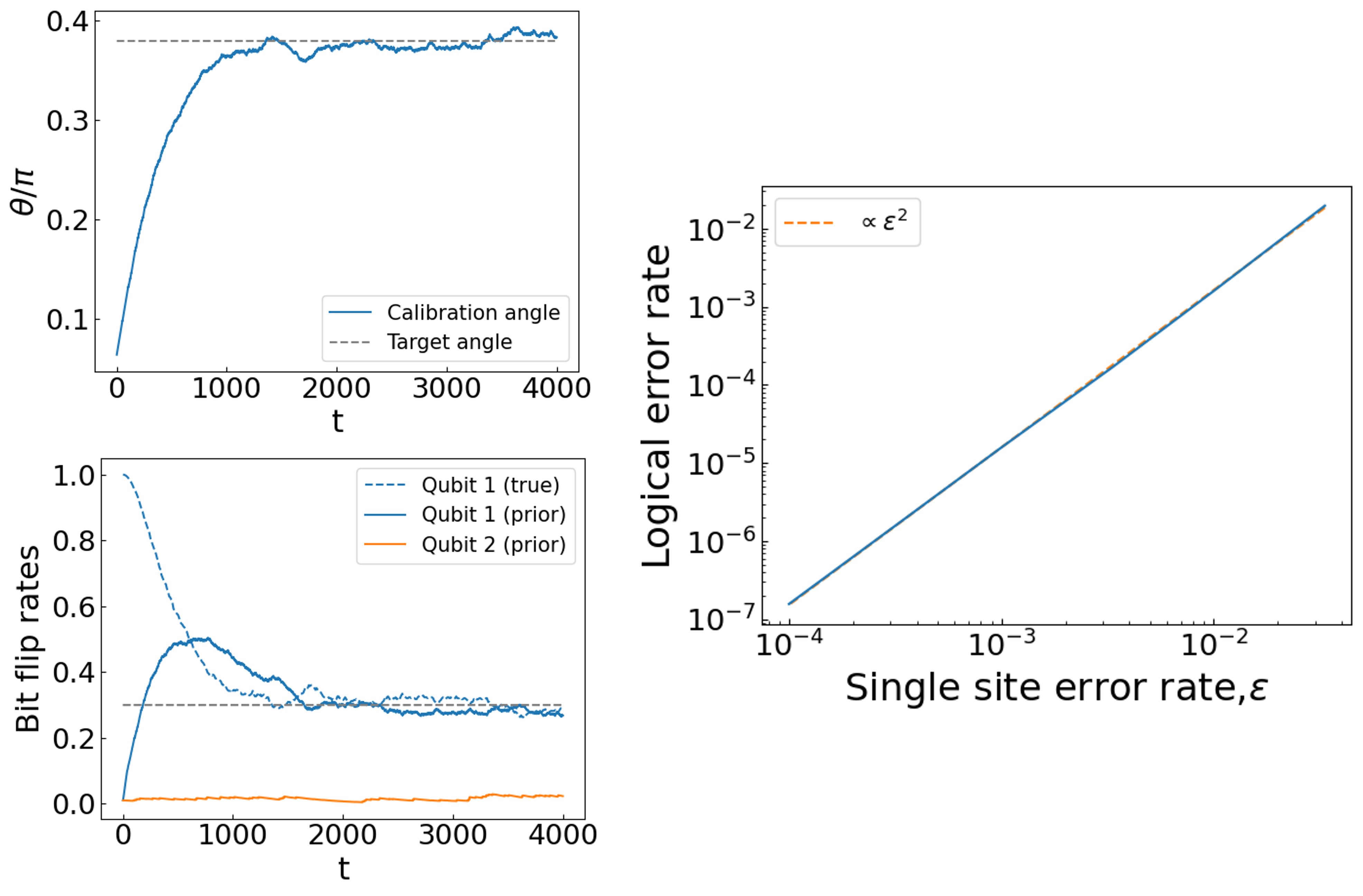}    
        \caption{Applying error correction with updated priors to calibration in the 
        unrotated surface code with distance 4. As explained in the main text, we switch to the unrotated surface code because it does not suffer from the degeneracy shown in Fig.~\ref{fig:Kalman}(top right) as it has no weight 2 stabilizers, leaving the rotated surface code to future work. (left top) The applied rotation angle $\theta$. Over the course of the experiment, it increases until  reaching  $\theta_{\rm target} - \theta_0$ (dashed line), and then oscillates. (left bottom) Estimated bit flip rates and the true bit flip rate for qubit 1 (dashed blue) over the course of the experiment. The actual bit flip rate from the gate $U_\theta$ decreases as a result of calibration feedback, while the prior is also updated, both ultimately stabilize around the target (dashed grey). (right) Decoder performance after calibration, fixing the actual and estimated bit flip rate of qubit 1 and varying the error rate $\varepsilon$ of all other qubits. Data agrees with  linear fit in log-log space (dashed orange) corresponding to $\varepsilon^2$ scaling.
        }
    \label{fig:calibration}
\end{figure}

Above we addressed the case where because of drift or other uncontrolled processes in qubit manufacture and control, we could correct errors in the presence of sites with high error rates with improved performance. Previously, these high-error sites were undesirable features of the hardware whose effect we sought to mitigate. We now investigate the complementary application, where using the adaptive decoder system allows us to protect against unknown errors while simultaneously extracting useful calibration data from a small number of physical qubits. For example, with a distance four code we can simultaneously provide calibration information for a single qubit gate and also detect and correct one unknown error. Specifically we examine the calibration of single-qubit gates alongside logical information storage, noting that a single qubit gate induces, from the perspective of a quantum error-correcting code, a weight-1 error of known location.

We consider the task of calibrating a rotation around the $X$-axis, $U_{\theta} = \exp(-i (\theta + \theta_0) X/2)$, where $\theta $ is an angle we can tune with control parameters in the lab, and $\theta_0$ is some unknown rotation offset (our `calibration error'). We have a desired rotation angle $\theta_{\rm target}$, and our goal is to set $\theta$ so that $\theta + \theta_0 \equiv \theta_{\rm tot} = \theta_{\rm target}$. Calibration and parameter estimation are closely related tasks. For the scenario we set up, correct calibration is equivalent to estimating $\theta_{\rm tot}$ and adjusting $\theta$ until it takes the desired value. 

One standard technique for this is a Ramsey-type protocol. Suppose one is characterizing a $\pi/2$ gate. Then a single round of the experiment has the steps (1) prepare the qubit in the state $\ket{0}$, (2) apply $U_{\theta = \pi/2}$, (3) measure in the computational basis, and if the calibration is correct, the final result is sampled from the uniform distribution. Using techniques similar to those of the previous section, the control parameter $\theta$ can be incremented or decremented in response to the measurement record so that it will keep $\theta + \theta_0 \approx \theta_{\rm target} = \pi/2$. In order to deploy this protocol in the error-correction context, we need a protocol for estimating the gate parameter $\theta_0$ alongside the preservation of logical information in the code.

We could, if we wished carry out the entire standard protocol without modification. The combined effect of measuring one qubit in the computational basis, applying $U_\theta$, and measuring in the computational basis again, assuming the ability to measure one qubit at a time, would resolve after measuring all stabilizers to a single qubit error on the known site, or effectively as an erasure error. However, we can carry out an analogous protocol even if single site measurement is hard. Our protocol instead is to start from an arbitrary logical state which has known stabilizer eigenvalues, apply $U_\theta$, and then measure the stabilizers. This induces a high probability error at the target location.

Concretely, we consider a gate on the first physical qubit, i.e. calibrating the angle of some single-qubit gate $U_\theta$ with \begin{equation}
    U_\theta = \exp(-i \frac{\theta}{2} X_1) = I \cos\frac{\theta}{2} - i X_1 \sin\frac{\theta}{2}.
\end{equation} Now suppose we have a state in the code space $\ket{\psi_L}$. We begin by applying $U_\theta$ and then measure the stabilizers. Suppose that there were no errors apart from the error induced by the calibration gate. Then the evolution of the density matrix $\rho_L \equiv \ket{\psi_L}\!\bra{\psi_L} $ through the protocol, including an average over stabilizer measurement outcomes, would be 
a sum of two terms
\begin{align}
    &\rho_L \to  U_{\theta} \rho_L U_{\theta}^\dag \to \sum_{s \in \mathbb{Z}_2^{k_z}} K_s U_{\theta} \rho_L U_{\theta}^\dag K^\dag_s  = \cos^2\frac{\theta}{2} \rho_L + \sin^2\frac{\theta}{2} X_1 \rho_L X_1 
   \end{align}
where the index $s$ in the POVM labels the possible syndromes $\{K^\dag_s K_s\}$, and in particular 
\begin{align}
    \sum_{s \in \mathbb{Z}_2^{k_Z}} K_s U_{\theta} \rho_L U_{\theta}^\dag K^\dag_s &= K_0 U_{\theta} \rho_L U_{\theta}^\dag K^\dag_0 + K_{X_1} U_{\theta} \rho_L U_{\theta}^\dag K^\dag_{X_1}\\
     K_0 U_{\theta} \rho_L U_{\theta}^\dag K^\dag_0 &= \cos^2\frac{\theta}{2} \rho_L\\
    K_{X_1} U_{\theta} \rho_L U_{\theta}^\dag K^\dag_{X_1} &= \sin^2\frac{\theta}{2} X_1 \rho_L X_1
\end{align}
with $K_0$ the projector onto the trivial syndrome subspace (i.e. the logical subspace) and $K_{X_1}$ the projector onto the subspace with the syndrome of an $X_1$ error (guaranteed to be orthogonal to the logical subspace for any code of nonzero distance). These two measurement outcomes have the probabilities $\cos^2 \frac{\theta}{2},\sin^2 \frac{\theta}{2}$ respectively. In this case the procedure for calibration is morally identical to the Ramsey protocol, expect that the probabilities $\cos^2 \frac{\theta}{2}, \sin^2 \frac{\theta}{2}$ are associated not with the measurement outcomes $\ket{0}$ and $\ket{1}$, but with the outcomes of trivial syndrome and nontrivial syndrome. 

Two things change when we allow for stochastic errors on qubits other than the target qubit. First, we must find some way of grouping the observed syndromes into those associated with bit flip errors supported on qubit 1, which we lump in with the pure $X_1$ error syndrome for purposes of calibration, and those associated with errors not supported on $X_1$, which we lump together with the trivial syndrome. Two phenomena make these assignments challenging. First, there is the fact that simultaneous errors on multiple sites near the qubit of interest may cause the overall error to be stabilizer equivalent to a lower weight error on some other qubit, and there is the related general fact that the errors which give rise to a particular syndrome subspace are not unique. We investigate the simplest possible approach to this problem: if the correction proposed by the decoder is supported on qubit 1 we ascribe this to the $X_1$ term in $U_\theta$, if the correction is not supported on qubit 1 we ascribe this to the identity component of $U_\theta$. Our expectation that this approach will work relies on the techniques developed in the previous sections, using decoders with prior information on the sites with high error rates so that the proposed corrections will prefer low-weight errors supported on the qubit of interest. However, it is ultimately an empirical question whether this approach is viable, and furthermore it is likely to only be effective in the regime of low background error rates. We view this technique not as a replacement for dedicated rounds of high-precision calibration, but as a means of extending the time for which the system remains well-controlled. Not also that this naive approach will fail on the rotated surface code for target qubits on the boundary, as the rotated surface code contains stabilizers of only weight 2. We anticipate that simple adjustments to the protocol are possible, but leave this to future work and use the unrotated surface code in the numerics that follow. 

We use a simple calibration update loop. The experimental parameter $\theta$ varies over rounds so we now index it with $\theta_t$ representing the value of $\theta$ during round $t$. The desired probability $p_{\rm target}$ associated with applying $X_1$ and the actual probability $p_t$ of applying $X_1$ for round $t$ are given by
\begin{equation}
     p_{\rm target} = \sin^2\frac{\theta_{\rm target}}{2}, \quad p_t =    \sin^2\frac{\theta_t+\theta_0}{2}.
\end{equation}
For a single run of experiment in which we prepare a logical state, apply $U_\theta$, measure the stabilizers, and finally run the BPOSD decoder to obtain a correction $c_t$, we create a single-shot estimate $b_t$ of $p_t$ as
\begin{equation}
    b_t = \begin{cases}
        1 & \text{$c_t$  supported on qubit 1}\\ 
        0 & \text{otherwise}
    \end{cases}.
\end{equation}

Then, defining the range of $\theta_{\rm target}$  as $\theta_{\rm target} \in ( -\pi,\pi )$, we update $\theta$ according to 
\begin{equation}\label{eq:calibration_update}
    \theta_{t+1} = \theta_t  -  \text{sgn}( \gamma_\theta) |\gamma_{\theta}| (  b_t - p_{\rm target}),  \quad   \text{sgn}( \gamma_\theta) \equiv \text{sgn}(\sin \frac{\theta_{\rm target}}{2})
\end{equation}
where, as in the Kalman case, $\gamma_\theta$ is a gain parameter whose magnitude ultimately is tunable, though with a reliable model for how the control is related to the action on the state, more sophisticated assignments of $\gamma_\theta$ derived from model parameters are possible. The assignment of a sign to $\gamma_\theta$ is simply a way to distinguish between $\pm \theta_{\rm target},$ as with $b_t \to p_t$, Eq.~\ref{eq:calibration_update} has a stable fixed point at $\theta_{\rm tot} = \theta_{\rm target}$ and an unstable fixed point at $\theta_{\rm tot} = -\theta_{\rm target}$. At the end of every round we also update the priors used in our BPOSD decoder as in the previous section.

The results of this protocol are shown in Fig.~\ref{fig:calibration}. We run our numerics for the unrotated surface code of distance 4. We do this because the unrotated code has minimum stabilizer weight 3, unlike the rotated version which contains weight 2 stabilizers. As we showed in Fig.~\ref{fig:Kalman}, the weight 2 stabilizer leads to an ambiguity between bit flips that occur on qubit 1 versus qubit 2, an ambiguity which does not affect successful decoding but which does matter for the simple calibration protocol we describe. While a more sophisticated use of decoding output could likely account for this stabilizer structure, to keep things conceptually simple for this work we instead use the unrotated surface code. This variant of the surface is distinct as every single-qubit error has a unique syndrome signature. As desired, as the experimental rounds proceed, both the control angle $\theta$ and the updated estimates of qubit-wise bit flip rates converge to their target values. The second part of Figure 4 verifies that $\varepsilon^2$ logical error rate scaling is maintained if we vary the background error rate for qubits besides qubit 1, showing we can run calibration while processing logical information with tolerable overhead. The parameters to make the performance plot are chosen to estimate the worst case performance after convergence (taken to be halfway through at $t = 2000$), by choosing the minimum estimated qubit 1 bit flip rate out of the fluctuating priors for $2000 < t < 4000 $. This is a worst case estimate because if the fluctuations were too large, we would end up in the case of the i.i.d. decoder with bad qubit of Section III.

The calibration approach we show here requires $b_i < 1/2$ for our chosen decoder. Thus, in practice, we require calibrating smaller rotation angle gates, e.g., $\pi/3$. This still allows calibration for typical under/over rotation scenarios due to either local oscillator drift or drive power fluctuations. 

In addition to single qubit calibration, our lemma shows we can also cover two qubit calibration generically with distance 5 codes while still correcting one unknown error. This is for the specific case of the CPHASE gate between two qubits $i$ and $j$, which implements $\exp(-i \phi Z_i Z_j)$. We speculate that for some codes (such as Bacon-Shor codes or their extensions in the surface code context) these types of gates are more amenable as the operation corresponds to an action on the gauge qubits, though we leave this for future work.

\section{Outlook}
In this work we showed that improving the quality of prior knowledge passed to the decoder can have important consequences. In the case of codes with a small number of high-error rate qubits it can change the overall performance of logical information storage, and it can also be applied to allow for calibration operations carried out simultaneously with the processing of logical information with no reduction in the number of unknown errors which can be successfully corrected. This work is an important contribution to the current race to eke out performance gains in pursuit of practical quantum advantage on near- to middle-term devices.

This work also suggests a number of fruitful opportunities for further work. This work has remained in the sparse defect regime with phenomenological noise. Though on the one hand our protocol suggests itself as a general purpose tool  to adapt existing quantum error codes to defective physical qubits with out sacrificing any more distance than is necessary, in practice we need to be able to handle and model errors not only in the data qubits, but also in measurement qubits. The most likely extension needed to achieve this is a hypergraph decoder. We also cannot handle qubits which are broken (experimentally inaccessible) rather than error-prone.

A second important issue is the extension of this work to degenerate quantum codes, that is, quantum codes with gauge degrees of freedom \cite{bacon-shor}. Here, there is an opportunity to carry out calibration operations in a way that introduces no additional errors on the logical subspace. There is a further opportunity here to bring the whole arsenal of the quantum learning literature to bear on the gauge qubit subspaces. 

We mentioned in Section III that the same number of unknown errors code be corrected for the surface code even if multiple error-prone locations were introduced, provided that they were distant from each other (i.e. were not connected by a minimal weight logical operator). It is an open question how this scaling works in general, and also open for investigation how this scaling works in general for quantum  qLDPCs. 

\begin{acknowledgments}
We thank Lorcan Conlon, Michael Gullans, and Victor Albert for helpful conversations. This work was supported by the National Science Foundation under the NQVL grant.

\end{acknowledgments}



\bibliography{main_bib}

\end{document}